# Scanning tunneling microscopy study of the possible topological surface states in BiTeCl

Y. J. Yan[1], M. Q. Ren[1], X. Liu[1], Z. C. Huang[1], J. Jiang[1], Q. Fan[1], J. Miao[1], B. P. Xie[1,2], T. Zhang[1,2*], D. L. Feng[1,2*]

[1] State Key Laboratory of Surface Physics, Department of Physics, and Advanced Materials Laboratory, Fudan University, Shanghai 200433, China
[2] Collaborative Innovation Center of Advanced Microstructures, Fudan University, Shanghai 200433, China

*E-mail of T. Zhang: tzhang18@fudan.edu.cn
*E-mail of D. L. Feng: dlfeng@fudan.edu.cn

**Recently, the non-centrosymmetric bismuth tellurohalides such as BiTeCl are being studied as possible candidates of topological insulators. While some photoemission studies showed that BiTeCl is an inversion asymmetric topological insulator, others showed that it is a normal semiconductor with Rashba splitting. Meanwhile, first-principle calculations failed to confirm the existence of topological surface states in BiTeCl so far. Therefore, the topological nature of BiTeCl requires further investigation. Here we report low temperature scanning tunneling microscopy study on the surface states of BiTeCl single crystals. On the tellurium-terminated surfaces with low defect density, strong evidences for topological surface states are found in the quasi-particle interference patterns generated by the scattering of these states, both in the anisotropy of the scattering vectors and the fast decay of the interference near step edges. Meanwhile, on samples with much higher defect densities, we observed surface states that behave differently. Our results help to resolve the current controversy on the topological nature of BiTeCl.**

Topological insulators (TIs) are bulk insulators with metallic surface (edge) states protected by time-reversal symmetry[1,2]. This intriguing state was theoretically predicted and experimentally observed in a variety of 2D and 3D systems with strong spin-orbital coupling (SOC)[3-7]. TIs bring forward new exotic physics such as magnetic monopole[8] and Majorana fermions[9]. So far, most of the discovered TIs are centrosymmetric. However, when inversion symmetry is absent in TIs, more unusual topological phenomena and practical applications can be realized, such as crystalline-surface dependent topological states, intrinsic topological p-n junctions[10,11], and topological magneto-electric effect[12]. Therefore, the search for inversion asymmetric TIs has recently attracted much interest.

The layered compounds BiTe$X$ ($X$ = Cl, Br, I) are of growing interest for their strong SOC and lack of lattice inversion symmetry. Large Rashba spin splitting (RSS) is widely observed in these compounds, which may facilitate future spintronics applications[13-16]. Moreover, a recent angle-resolved photoemission spectroscopy (ARPES) study revealed that BiTeCl cleaved in vacuum exhibited surface states with Dirac cone like dispersion, implying that it is possibly a topological insulator without inversion symmetry[17]. However, the emergence of topological phase in BiTeCl was not found in first-principles calculations[15], and some other ARPES works on BiTeCl only observed parabolic bands with RSS, instead of a Dirac cone[16]. Therefore, whether the surface state in BiTeCl has a topological origin and how the unexpected TI phase emerges are still open issues, which require further investigations.

Scanning tunneling microscopy (STM) is a surface sensitive tool which played important role in the discovery of TIs[18-20] (Surface p-n junction was observed on BiTeI by STM recently [21, 22]). In this paper, we report first STM study on the surface states of BiTeCl. On the sample with relatively low defect density, we observed surface state induced interference patterns on the Te-terminated surface. The scattering intensity is concentrated along the $\overline{\Gamma}-\overline{M}$ direction, which resembles the classical topological insulator $Bi_2Te_3$. Moreover, the decay of the standing waves around step edges is much faster than those of conventional surface states. By comparing with previous ARPES results and simulated scattering patterns, the observed surface states are most likely of topological origin. More interestingly, on the sample with higher (about 5 times) defect density, we observed surface states with different bands and scattering patterns. It implies that a band structure mutation may happen with increased bulk carriers, which will screen out the surface polarization.

## Results and discussion

**Sample characterization and surface topography.** BiTeCl single crystal preparation and STM experimental details are described in the Methods section. The crystal structure of BiTeCl is shown in Fig. 1a, characterized by the alternative stacking of Te, Bi and Cl layers along the *c*-axis. Each atomic layer is arranged in a triangular lattice. This crystal structure is confirmed by x-ray diffraction (XRD) (Fig.1b) with the lattice constants c = 12.39 Å. The actual Bi, Te, and Cl ratios were determined as 1:1.1:0.91 by energy dispersive spectroscopy. Temperature dependence of the resistivity is shown in Fig. 1c, which displays a metallic behavior, similar to that reported before[16].

It is known that the cleavage of BiTeCl happens between the Te and Cl layers, leaving a pair of polarized Te- and Cl-terminated surfaces[17]. In our STM measurement, two kinds of topographies corresponding to the Te- and Cl-terminated surfaces are observed, as shown in Figs. 1d,e. The Te-terminated surface is atomically flat with several kinds of point defects, as marked by I, II and III. The type I defects are most common, and they are clover-shaped at Bias voltage $V_b > 0.5$ V. Fig .1f is the Te lattice imaged at $V_b = 30$ mV with the in-plane constant of 0.42 nm. At this low $V_b$, type I defects appear like triangular dips, as marked at the top-left corner in Fig. 1f. The center of the triangle is at the hollow site of the Te lattice, thus type I defect is possibly caused by substituting a Te atom at the Bi site ($Te_{Bi}$). $Te_{Bi}$ defects are electron donors, which is consistent with the n-type nature of bulk BiTeCl[16, 23]. The averaged defects density (including all kinds of defects) for the surface shown in Fig. 1d is about 2~3×10$^{12}$ cm$^{-2}$. If all the observed defects belong to the topmost Te-Bi-Cl tri-layer, the corresponding bulk defect density will be 3~5 ×10$^{19}$ cm$^{-3}$. Samples with higher defect densities are also studied, which will be discussed later. While for the Cl-terminated surface, the topography is much rougher with random distributed clusters and some stripe like features. This could be due to the high reactivity of Cl-ions and stronger polarity of the surface. Fig. 1g displays the tunneling spectroscopies (dI/dV) taken on the two surfaces, which gives a measurement of local density of states (LDOS). For the Te-terminated surface, a rather V-shaped LDOS is observed in between -200 ~ -700 meV, with a minimum at ~ -600 meV. This feature is consistent with the existence of Dirac cone[6,19]. The possible onsets of bulk conduction and valence bands are indicated by the arrows in Fig. 1g. As for the Cl-terminated surface, the dI/dV shows a gap like structure from -400 meV to 100 meV. Although surface states were predicted on the ideal Cl-terminated surface[15], practically this is not observed here. Below we will focus on the Te-terminated surfaces.

**QPI measurements on Te-terminated surface.** To further study the electronic states, we performed dI/dV mappings on the Te-terminated surface. Figs. 2a–f show representative dI/dV maps at various $V_b$ measured in the same region shown in Fig. 1d. A marked feature is that pronounced interference patterns (or standing waves) can be observed in the vicinity of the defects. It is known that such a quasi-particle interference (QPI) pattern is caused by the scattering between initial and final states at the same constant energy contour (CEC) in the momentum space. The corresponding wavevectors are $k_i$ and $k_f$, respectively, and the scattering vector is given by $q = k_f - k_i$. Pronounced interference usually comes from 2D states because the bulk (3D) states would give continuous ranges of $q$ on the projected surface Brillouin zone (SBZ) and cannot generate distinct interference. In Figs. 2g-l, we show the 2D Fourier transformation (FT) of Figs. 2a-f, in which the scattering intensities with various $q_s$ can be revealed (the 2D FTs are six-fold symmetrized to raise signal to noise ratio). SBZ is superposed in Fig. 2g to help orient the scattering directions. One finds that the $q$ vectors make up an anisotropic ring, the size of which shrinks as $V_b$ decreases, indicating that the scatterings are from a dispersive electron like surface band. The regions with high scattering intensity are along the $\overline{\Gamma} - \overline{M}$ directions. Fig. 2m shows the line cuts that are extracted along $\overline{\Gamma} - \overline{M}$ in all of the FT maps (displayed in a gray scale). The $E$ vs. $q$ dispersion can be clearly visualized. Further deciphering the FT patterns requires detailed knowledge of the band structure. As mentioned earlier, ARPES studies have observed two kinds of surface bands in BiTeCl, which are: (i) Dirac cone like surface band with the Dirac point (DP) at ≈ -560 meV [17]; (ii) Rashba split parabolic bands with the band bottom located at ≈ -450 meV[16] (note that ref. 17 also observed parabolic bands in some cases which they ascribed to single layer BiTeCl flakes). Below we will distinguish which one is actually observed here.

The topological Dirac cone and Rashba split surface bands both have spin textures on their CEC. For scatterings that preserve time-reversal symmetry, the intensity of a specific $q$ can be calculated by spin-dependent joint density of states (SJD):

$$SJD(q) = \int T(k,q)I(k)I(k+q)d^2k .$$

*I(k)* is the intensity of an electronic state with *k* at certain CEC, as directly measured by ARPES. $T(k,q) = |\langle S(k)|S(k+q)\rangle|^2$ is the matrix element considering the spin states of initial and final state. At the energy range where we observed interference patterns (-300 ~ 100 meV), the CECs of both cases (i) and (ii) deviate from a circular shape, due to the warping effect (the high order term of SOC that reflects the surface crystal symmetry)[24, 25]. We calculated the SJD corresponding to these two cases. For case (i), the CEC actually resembles classical topological insulator $Bi_2Te_3$, in which the band structure can be described by the effective Hamiltonian with a warping term[24]:

$$H(k) = E_0 + v_F(k_x\sigma_y - k_y\sigma_x) + \frac{\lambda}{2}(k_+^3 + k_-^3)\sigma_z$$

Fig. 2n shows the simulated CEC at *E* = -50 meV to reproduce the ARPES result in ref. 17 (Here we have chosen $E_0$ = -0.56 eV, $v_F$ =2.65 eV Å, and $\lambda$ = 60 eV Å$^3$). The CEC is a warped hexagon, as shown in ref. 17, the valley points (along $\overline{\Gamma}-\overline{K}$) are better separated from the bulk bands than the "tip" points (along $\overline{\Gamma}-\overline{M}$). The scattering vectors connecting valley points are marked as $q_1$~$q_3$. It has been demonstrated in $Bi_2Te_3$ that due to the spin chirality of the Dirac cone, scatterings at $q_1$ and $q_3$ are suppressed for their opposite (or nearly opposite) spin orientations. Thus the dominant scattering vector is $q_2$, which is along the $\overline{\Gamma}-\overline{M}$ directions[19]. The simulated SJD pattern in Fig. 2o also gives high intensity direction along $\overline{\Gamma}-\overline{M}$, qualitatively agrees with the data. A full T-matrix calculation of the scattering for this case is reported in ref. 26, which gives a similar FT pattern. For case (ii), the CEC is composed of two (inner and outer) Rashba bands, which can be described by the following effective Hamiltonian proposed in ref. 25:

$$H(k) = E_0 + \frac{k_{//}^2}{2m^*} + v_c(1 + a_c k_{//}^2)(k_x\sigma_y - k_y\sigma_x) + \lambda(3k_x^2 - k_y^2)k_y\sigma_z$$

Fig. 2p shows the CEC at E = 0 meV to reproduce the Rashba bands observed in ref. 16. (We have chosen m* = 22.1 eV Å$^2$, $E_0$ = -0.52 eV, $v_c$ =1.15 eV Å, $a_c$ = -41 Å$^2$, $\lambda$= 30 eV Å$^3$). Here because inner and outer sub-bands have opposite spin chirality, the dominant scattering vectors will be the ones connecting two sub-bands (marked as $q_4$ and $q_5$). However, although the CEC has anisotropy, the simulated SJD pattern (Fig. 2q) does not show significant enhancement along $\overline{\Gamma}-\overline{M}$. So based upon the above simulations, the anisotropic interference supports that the surface states here have a topologically nontrivial origin, rather than a normal 2D states with Rashba splitting.

As shown in Fig. 2n, $q_2$ is about $\sqrt{3}$ times of the *k* along $\overline{\Gamma}-\overline{M}$ ($k_{\Gamma M}$). Thus we can extract the band dispersion E - $k_{\Gamma M}$ from Fig. 2m. The results are shown in Fig. 3k. A linear fitting gives a slope ($v_F$) of 2.34 eV Å and an intercept of $E_d$ = -580 meV (which is the location of Dirac point). Both $v_F$ and $E_d$ qualitatively agree with the topological surface states observed by ARPES in ref. 17.

**QPI measurements along atomic step edges.** Besides point defects, atomic steps on the cleaved surface can also induce scattering that contains information on the topological nature of the surface states. To further check the above findings, we measured interference around the step edges. Fig. 3a shows a step edge formed along $[10\bar{1}]$ direction. We mapped an area next to the upper step edge that has no point defects (as marked in Fig. 3a). The dI/dV mappings in Figs. 3b-h show that a standing wave clearly forms around the edge, but decays quickly away from the edge. For a straight step, the scattering is reduced to a 1D problem with $q$ perpendicular to the edge. For conventional surface states, it is known that the decay of the standing wave intensity follows the power-law: $I = \sin(qx+\varphi)x^p$, with an index $p$ of -1/2 (ref 27). But as recently discovered, if $q$ is a vector which connects opposite spin states (for example, the backscattering vector of topological surface states), the decay will be faster with $p$ = -3/2 (ref 28). Thus the decay behavior will give another clue to distinguish topological surface states. In our case, the $q$ is along $[11\bar{2}]$, which is the $\bar{\Gamma}-\bar{K}$ direction in the $k$ spaces. For the CEC displayed in Fig. 2n, the dominant $q$ along $\bar{\Gamma}-\bar{K}$ is $q_1$, which is exactly the backscattering vector that connects opposite spin states. For the CEC shown in Fig. 2p, $q_4$ is the backscattering vector along $\bar{\Gamma}-\bar{K}$ but connects the states with the same spin orientation. So the decay in case (i) will follow $p$ = -3/2, and for case (ii) the decay will be the same as conventional surface states with $p$ = -1/2. Figs. 3i, j show two representative fittings of the line profiles extracted from dI/dV maps, corresponding to $V_b$ = -250 meV and -50 meV. Both of them can be fitted with a decay index of $p$ = -3/2. Thus this is the second evidence to substantiate that we have observed topological surface states. Because $q_1$ = $2k_{\Gamma M}$, we can also extract $E - k_{\Gamma M}$ relation from the mappings near a step edge, as shown in Fig. 3k. The $E - k_{\Gamma M}$ obtained by two different ways basically match with each other.

**QPI measurements on Te-terminated surface with higher defects density.** If BiTeCl is a topological insulator, one question is why it was not reproduced in previous theoretical calculations. As discussed in ref. 17, a possible reason is that the vacuum-cleaved surface is polarized (unpassivated), which will induce electrical field in the sample. The electrical field and the effective pressure generated by the field may significantly alter the bulk band structure. A topological transition is thus possible under these conditions[25]. A recent transport measurement also reports a possible topological phase transition of BiTeCl under high pressure[29]. In our measurement, the very clean Te-terminated surface does support the polarization assumption. On the other hand, the point defects (as shown in Fig. 1d) should provide bulk carriers, which will weaken the surface polarization induced intrinsic electrical field through screening. Thus one may expect that if there are sufficient defects, the topological surface states might vanish due to the change of bulk band structure. So far, we have presented data taken on the surface with defect density of $2\sim3\times10^{12}$ cm$^{-2}$. Fig. 4a shows a Te-terminated surface with much higher defect density of about

$1.2\times10^{13}$ cm$^{-2}$. The most common defects are still clover-shaped ones. The dI/dV maps on this surface and their FTs are shown in Figs. 4e-i and Figs. 4j-n, respectively. The line cuts taken along $\overline{\Gamma}-\overline{M}$ of the FTs are summarized in Fig. 4c. One finds that the interference patterns also exist, their FTs are less anisotropic and more blurred when compared with those in Fig. 2, but the *E-q* dispersion is still visible in Fig. 4c. As shown in Fig. 4d, the *E–q* relation extracted from Fig. 4c is clearly different from the *E-$q_2$* relation extracted from Fig. 2m. At first, Fig. 4d appears to indicate a rigid band shift ($E_F$ shift) between the two surfaces, that is, the $E_F$ of sample with higher defect density shifts downwards. However, we note that the bulk BiTeCl is n-type, more defects should shift $E_F$ upwards. This discrepancy, as well as the more isotropic scattering patterns, imply that there might be a band structure mutation instead of a rigid band shift. In Fig. 4b, we show the tunneling spectrum measured on this surface. It is significantly different from the V-shaped spectrum measured on the surface in Fig. 1d. In this regard, we propose that the surface states in Figs. 4e-i may be attributed to normal surface states, instead of a topological Dirac cone. In Fig. 4d, we show that a parabolic fitting of the E-q gives a band bottom at -310 meV. However, because the high density of defects also affect the resolution of the QPI, direct comparisons between the FTs and simulations (such as the one in Fig. 2q) are difficult here. More experimental and theoretical works are needed to understand the effect of defects.

In summary, we have studied the surface states on the Te-terminated surface of BiTeCl by low temperature STM. On the surface with relatively low defects density, the anisotropic scattering patterns and faster decay of the standing waves along atomic step edges strongly suggest that the surface states have a topological nontrivial origin. On the other hand, surface states observed on samples with higher defect densities might be attributed to the topological trivial states observed in some ARPES measurements. Our results help to resolve the current discrepancy in the ARPES studies of BiTeCl, and also suggest the possible causes for the emergence and disappearance of topological surface states, thus a possible path to tune the topological transition in such materials.

## Methods

**Crystal growth.** Single crystals of BiTeCl were grown by a self-flux method as described in Ref. 23. The precursor of $Bi_2Te_3$ polycrystal was obtained by heating Bi and Te powders with stoichiometric ratio at 823 K for 24 hours. Then the mixture of $Bi_2Te_3$ and $BiCl_3$ with the ratio of 1:9 was grounded and loaded into an evacuated quartz tube. All the sample processing procedures were performed in a glovebox filled with argon. The mixture was heated to 793 K, kept for 48 hours, and slowly cooled to 473 K. Single crystals of BiTeCl with size of several mm$^2$ is obtained through this procedure.

**Crystal cleavage.** Samples were mounted onto the sample holder by Ag epoxy. After transferred into the STM chamber with the vacuum $< 1\times10^{-10}$ *torr* and pre-cooled at 77 K, the single crystals were cleaved quickly by knocking off a small post attached to the samples and immediately transferred to STM module at the temperature of 4.5

K.

**STM measurements.** All the STM experiments were performed in a commercial cryogenic Createc STM system at 4.5 K. Normal Pt tips were used after treated on Au (111) surface. Bias voltage ($V_b$) is applied to the sample with respect to the tip, and tunneling spectroscopy (dI/dV) is collected using lock-in method with a modulation frequency of 975 Hz.

**Acknowledgements**

We thank professor Xin-Gao Gong for useful discussions. This work is supported by the National Science Foundation of China, and National Basic Research Program of China (973 Program) under the grant No. 2012CB921402, No. 2011CBA00112, and No. 2011CB921802.

**Author contributions**

The low-temperature STM/STS measurements were performed by Y. J. Yan and M. Q. Ren. The samples were prepared by Y. J. Yan and Z. C. Huang. The data processing was performed by Y. J. Yan, M. Q. Ren, X. Liu and T. Zhang. The theoretical simulation was performed by T. Zhang. Y. J. Yan and T. Zhang coordinated the whole work and wrote the manuscript. All authors have discussed the results and the interpretation.

**Additional information**

Competing financial interests: The authors declare no competing financial interests.

**Figure 1 | Characterizations of BiTeCl single crystals and topographic STM images.** Crystal structure (**a**), XRD (**b**), and resistivity (**c**) of BiTeCl. Typical topographic images on the Te-terminated (**d**) and Cl-terminated surfaces (**e**) of BiTeCl. Three types (I, II, III) of defects are marked. (**f**) Atomically resolved image of the Te-terminated surfaces. Dashed triangle shows the position of a type I defect. (**g**) Averaged dI/dV spectra taken on Te-and Cl- terminated surfaces.

**Figure 2 | QPI patterns on the Te-terminated surfaces.** (**a-f**) dI/dV maps taken in the area shown in Fig. 1**d** with various $V_b$. Each map has 250 × 250 pixels. (**g-l**) The Fourier transformations of **a-f**. (**m**) The line cuts extracted from the FTs along $\overline{\Gamma} - \overline{M}$, shown in gray scale. (**n**) The CEC of a topological surface state with warping effect (reproduced from ref. 17) and (**o**) calculated FT patterns corresponding to this CEC. (**p**) The CEC with inner- and outer- Rashba bands of a normal surface state with warping effect (reproduced from ref. 16), and (**q**) calculated FT patterns corresponding this CEC.

**Figure 3 | Decay behavior of the standing waves around step edges.** (**a**) An atomic step edge on Te-terminated surfaces of BiTeCl, formed along $[10\overline{1}]$ direction. The dashed rectangle shows the area for dI/dV mapping. (**b-h**) dI/dV maps taken at various $V_b$ which show standing waves formed along step edge. (**i**, **j**) Fitting of the intensities of standing wave mapped at $V_b$ = -250 meV and -50 meV, with the distance to step edge. (**k**) *E-k* relation extracted from Fig. 2**m** and Fig. 3.

**Figure 4 | QPI patterns on the Te-terminated surfaces with higher defects density.** (**a**) Topographic image of the Te-terminated surfaces with surface defects density of ~1.2×10$^{13}$ cm$^{-2}$. (**b**) Averaged dI/dV spectrum taken on Te- terminated surface with high defect density (red), with comparing to the one measured on the surface in Fig. 1d (black). (**e-i**) The dI /dV maps at various $V_b$. Each map has 250 × 250 pixels. (**j-n**) The FTs of the dI /dV maps in (**e-i**). (**c**) The line cuts extracted from the FTs along $\overline{\Gamma} - \overline{M}$, shown in gray color scale. (**d**) Red spots: *E-q* relation extracted from panel **c** here. Gray spots: *E-q* relation extracted from Fig. 2**m**.

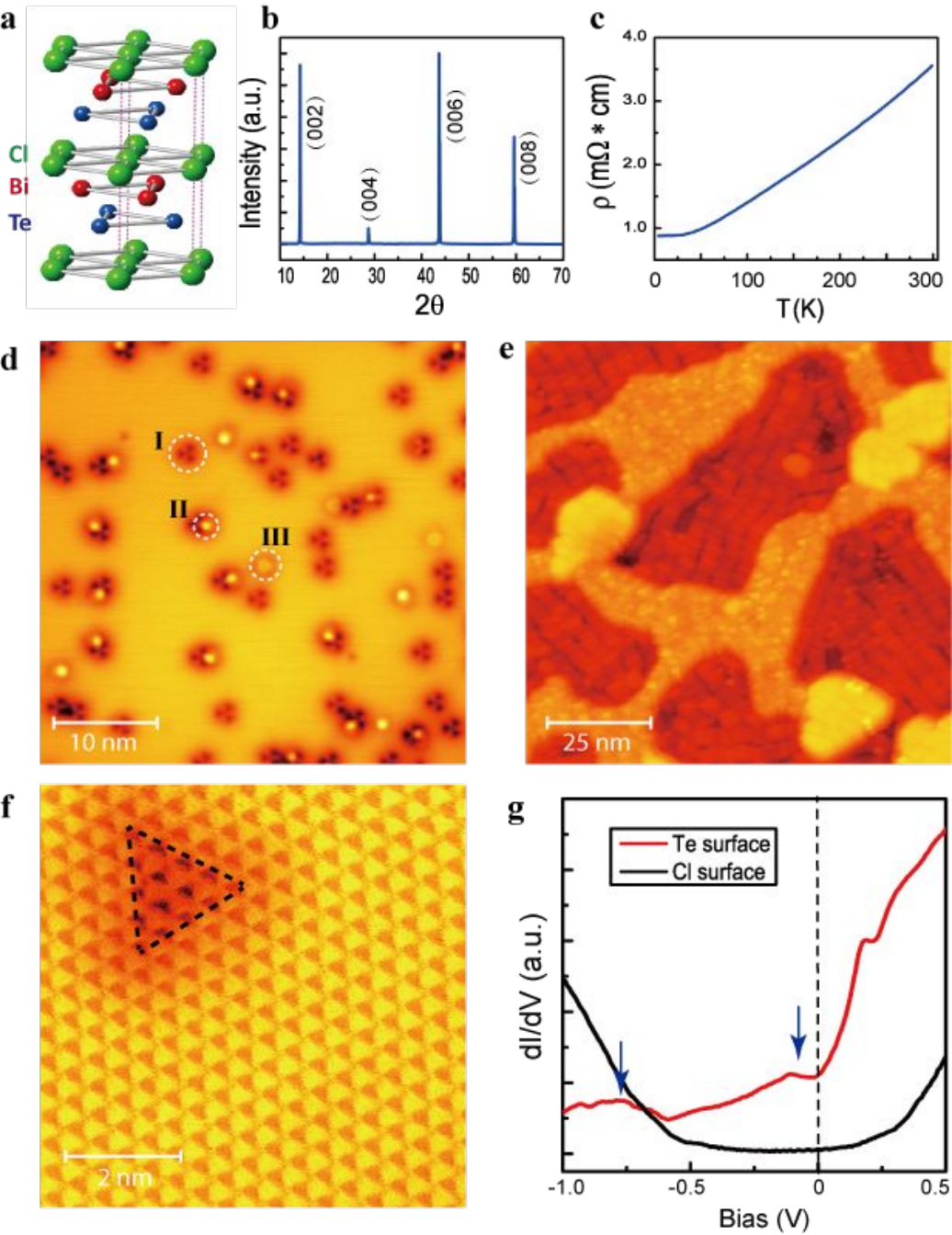

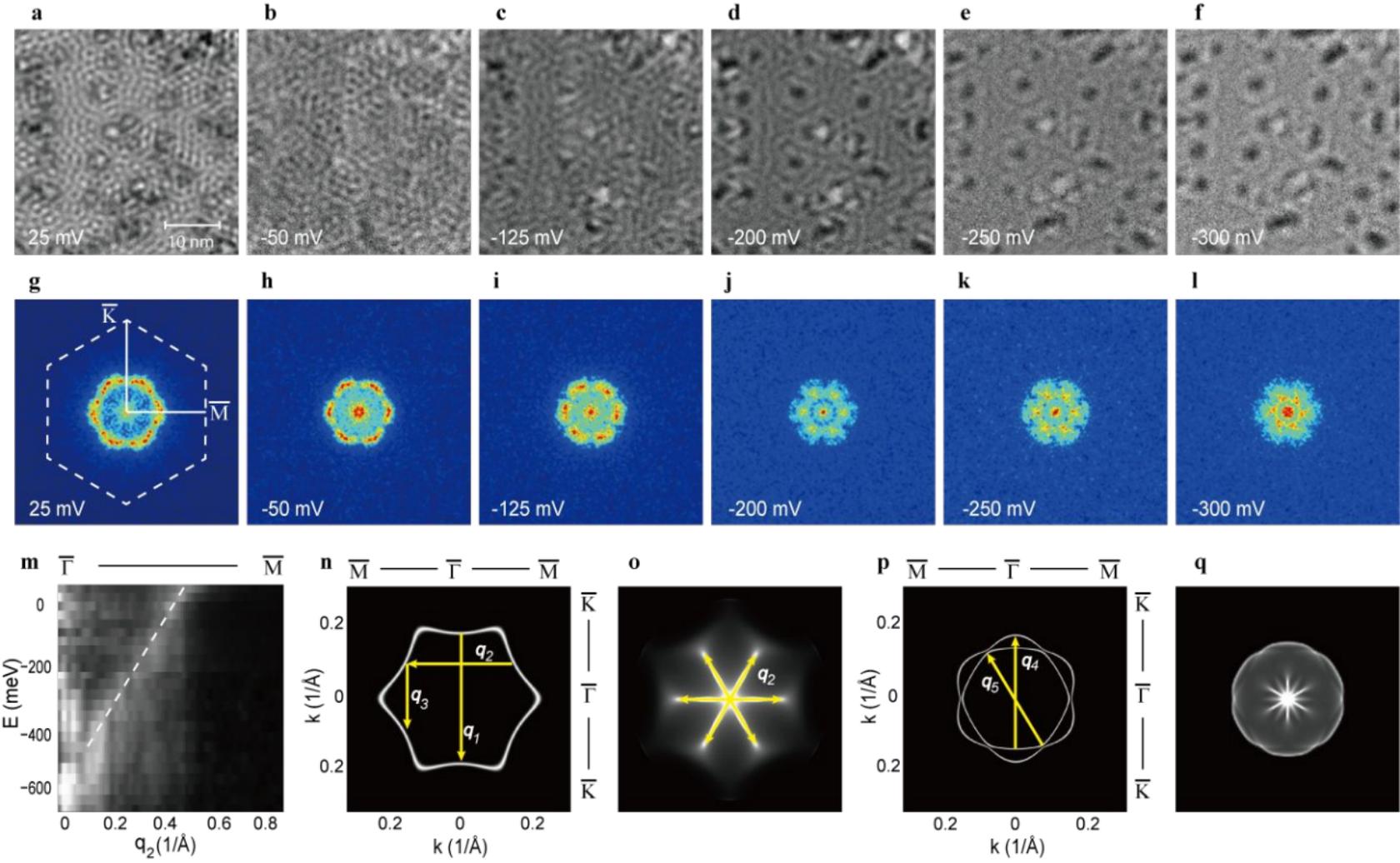

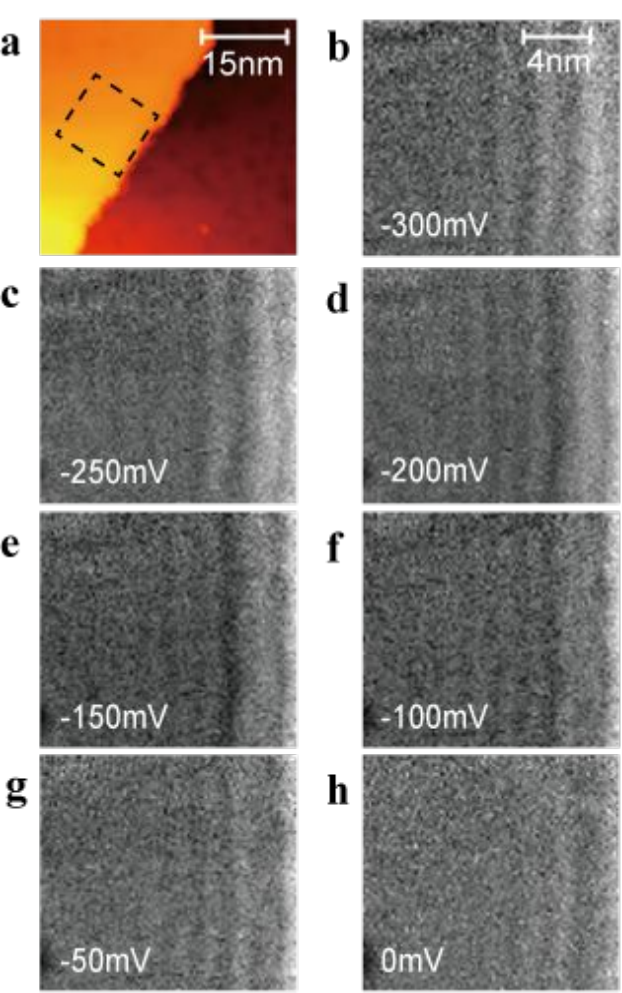

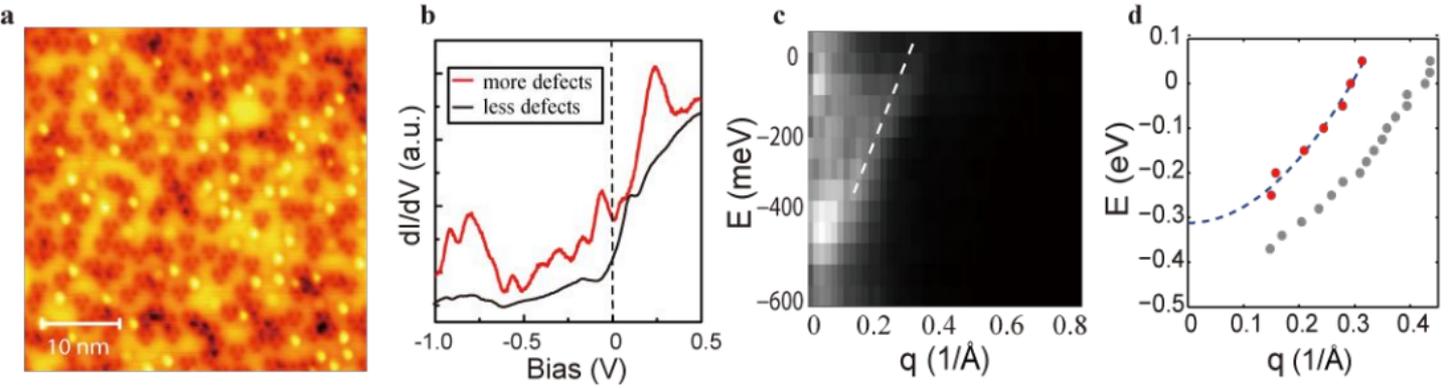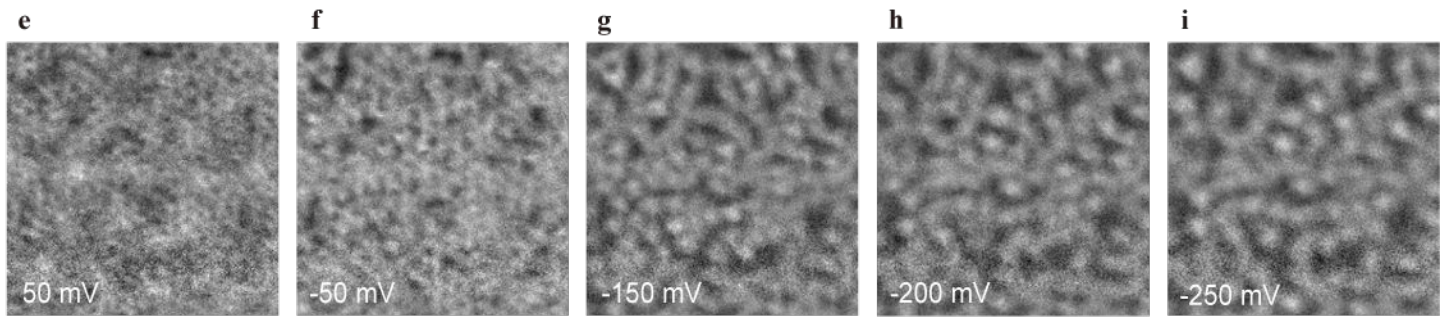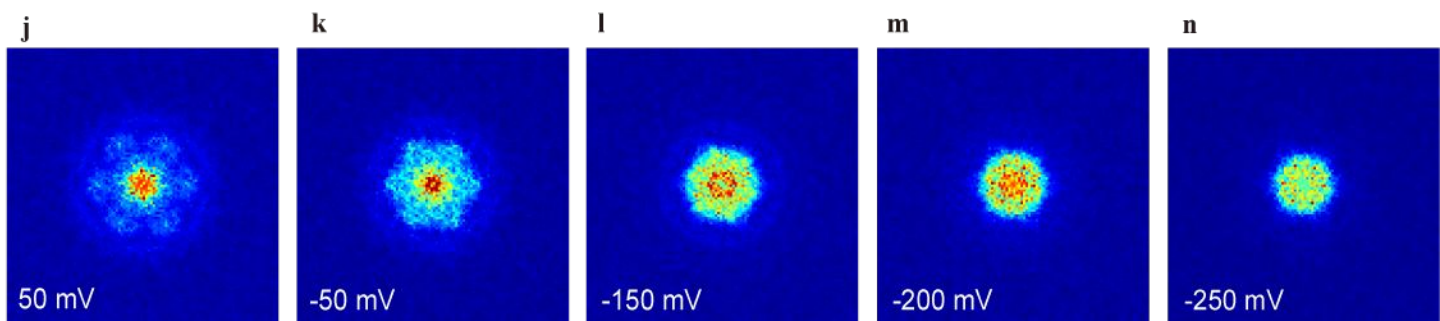